\documentclass[sigconf]{acmart}

\usepackage{color}
\usepackage{graphicx}   
\usepackage{array}
\usepackage{dcolumn}    
\usepackage{bm}         
\usepackage[normalem]{ulem}
\usepackage{amsmath}
\usepackage{soul}
\usepackage[nolist,nohyperlinks]{acronym}
\usepackage{upgreek}
\usepackage{subfigure}
\usepackage{epsfig}
\usepackage{caption}

\def\BibTeX{{\rm B\kern-.05em{\sc i\kern-.025em b}\kern-.08emT\kern-.1667em\lower.7ex\hbox{E}\kern-.125emX}}
    
\copyrightyear{2019} \acmYear{2019} \acmConference[NANOCOM '19]{The Sixth Annual ACM International Conference on Nanoscale Computing and Communication}{September 25--27, 2019}{Dublin, Ireland} \acmBooktitle{The Sixth Annual ACM International Conference on Nanoscale Computing and Communication (NANOCOM '19), September 25--27, 2019, Dublin, Ireland} \acmPrice{15.00} \acmDOI{10.1145/3345312.3345470} \acmISBN{978-1-4503-6897-1/19/09}

\begin{document}

%
\title{Workload Characterization of Programmable Metasurfaces}

%
\author{Taqua Saeed}
\authornote{Corresponding email: st009698@stud.fit.frederick.ac.cy}
\affiliation{%
  \institution{Frederick University Cyprus\\ \emph{and} University of Cyprus}
  \city{Nicosia}
  \country{Cyprus}
}

\author{Sergi Abadal}
\affiliation{%
  \institution{Universitat Polit\`{e}cnica de Catalunya}
  \city{Barcelona}
  \city{Spain}
}

\author{Christos Liaskos}
\affiliation{%
  \institution{Foundation of Research and Technology (FORTH)}
  \city{Heraklion}
 \country{Greece}
}

\author{Andreas Pitsillides}
\affiliation{%
  \institution{University of Cyprus}
  \city{Nicosia}
 \country{Cyprus}
}

\author{Hamidreza Taghvaee}
\affiliation{%
  \institution{Universitat Polit\`{e}cnica de Catalunya}
  \city{Barcelona}
  \city{Spain}
}

\author{Albert Cabellos-Aparicio}
\affiliation{%
  \institution{Universitat Polit\`{e}cnica de Catalunya}
  \city{Barcelona}
  \city{Spain}
}

\author{Marios Lestas}
\affiliation{%
  \institution{Frederick University Cyprus}
  \city{Nicosia}
  \country{Cyprus}
}

\author{Eduard Alarc\'{o}n}
\affiliation{%
  \institution{Universitat Polit\`{e}cnica de Catalunya}
  \city{Barcelona}
  \city{Spain}
}

%
\renewcommand{\shortauthors}{T. Saeed et al.}

%
\begin{abstract}
Metasurfaces are envisaged to play a key role in next-generation wireless systems due to their powerful control over electromagnetic waves. The last decade has witnessed huge advances in this regard, shifting from static to programmable metasurfaces. The HyperSurface (HSF) paradigm takes one step further by integrating a network of controllers within the device with the aim of adding intelligence, connectivity, and autonomy. However, little is known about the traffic that this network will have to support as the target electromagnetic function or boundary conditions change. In this paper, we lay down the foundations of a methodology to characterize the workload of programmable metasurfaces and then employ it to analyze the case of beam steering HSFs. We observe that traffic is bursty and highly dependent on the position of the target. These results will enable the early-stage evaluation of intra-HSF networks, as well as the estimation of the system performance and cost.
\end{abstract}

%
%


%

\maketitle

\section{Introduction}
Metasurfaces have garnered increasing attention in recent years due to the powerful control of electromagnetic waves that they have been proven to achieve \cite{Pendry2006}. Generally composed of a periodic array of subwavelength metallic inclusions over a substrate (see Fig. 1a), metasurfaces allow to implement functionalities such as anomalous reflection, absorption, polarization control, or focusing \cite{Glybovski2016, Li2018}. A myriad of designs have been appearing that exploit such potential for a wide range of frequencies in applications like electromagnetic cloaking, beam steering, or holographic displays \cite{Vellucci2017, Tasolamprou2017, Tsilipakos2018a, ChenXZ2012, Li2017b}.  

Two of the main downturns of most metasurfaces are non-adaptivity and non-reconfigurability as, in most cases, the electromagnetic function and its scope are fixed once the unit cell is designed. In response to these drawbacks, metasurface design with tunable or switchable elements have emerged \cite{Oliveri2015}. The resulting reconfigurable metasurfaces can be globally or locally tunable and, with the appropriate control means, they become programmable \cite{Liu2018ISCAS}. In fact, the potential of the combination of local tunability and digital control (via FPGAs, as shown in Fig. 1b) has been showcased in several prototypes of programmable metasurfaces for polarization control, focusing, or beam manipulation \cite{Cui2014, Hosseininejad2019, Li2017b}. 

One step further towards general-purpose, autonomous, intelligent metasurfaces is the HyperSurface (HSF) paradigm \cite{Liaskos:2018:UAS:3289258.3192336,AbadalACCESS}. As shown in Fig. 1c and further detailed in Sec. \ref{sec:background}, HSFs integrate a network of controllers within the structure of the metasurface. Controllers drive the reconfigurable unit cells and exchange information with neighbouring controllers so that the HSF can (i) implement a given electromagnetic functionality requested by an authorized user, and (ii) adapt to changes in the environment. The internal network of controllers is the enabler of the HSF approach and the main difference with respect to conventional programmable metasurfaces. This work focuses on this aspect.

\begin{figure}[!t]
\includegraphics[width=\columnwidth]{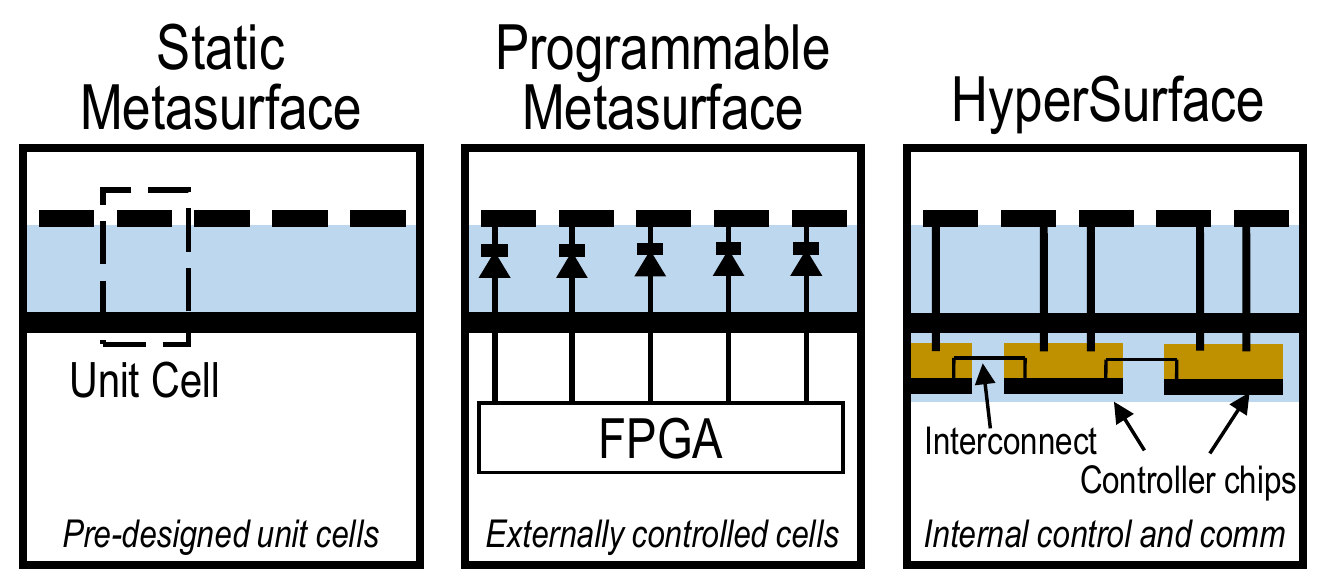} 
\vspace{-0.7cm}
\caption{Progress in metasurface design from static to programmable with integrated tuning and control electronics.}
\vspace{-0.5cm}
\label{fig:intro}
\end{figure}

To approach the design of the internal network of controllers properly, it is necessary to understand the workload that the internal HSF network will have to serve. In other words, we need to characterize the expected traffic between the HSF's controllers in order to take informed design decisions regarding the HSF network. Due to the novelty of the programmable metasurface in general and of the HSF concept in particular, such an analysis has not been carried out thus far. 

This paper aims to bridge this gap by providing a traffic analysis of programmable metasurfaces. Given the breadth of the problem, we first focus on a particular yet relevant functionality: anomalous reflection for beam steering. To this end, we develop a methodology to obtain the state of each unit cell as a function of several design parameters (e.g. unit cell size) and application parameters (e.g. steering angle). We apply the methodology to study the variations in the unit cells' states as functions of changes in the required reflection angle. Assuming that every state change implies sending a message, these variations provide spatiotemporal information on the metasurface traffic. By mapping angular changes to physical movements in real-world scenarios, our methodology can be applied to real use cases such as tracking a mobile user in a 5G network.

The remainder of this paper is organized as follows. Sec. \ref{sec:background} describes the HSF paradigm. Sec. \ref{sec:programming} outlines the system model, along with the programming rules for anomalous reflection. Sec. \ref{sec:methodology} lays down the evaluation methodology. Then, Sec. \ref{sec:results} presents the workload characterization results and Sec. \ref{sec:conc} concludes the paper.

\section{\label{sec:background}Hypersurfaces: Toward Autonomous Intelligent Metasurfaces}

Metasurfaces transform impinging EM waves into a response that
fulfills some user-specified form. Common examples include steering
a planar wave to any direction, while altering its power, polarization
and phase in any joint (or disjoint) manner~\cite{Li2018}.
To attain this objective, metasurfaces rely on the Huygens principle,
which states that any EM wavefront can be traced back to a planar
current distribution that generates it~\cite{Pao.1976}. Thus, the
active elements within each meta-atom essentially force the surface
currents created by the impinging EM wave into following a specific
distribution that corresponds to the user-specified EM response. Towards
this end, the metasurface needs to account for the EM interactions
between meta-atoms, i.e., various inductive currents created due to
the cross-meta-atom effects~\cite{Cui.2017}.

The HyperSurface paradigm dictates a framework for: i) Describing
and understanding the physics behind metasurfaces in computer networking
and software terms. ii) Enabling communication capabilities between
metasurfaces and any other smart device via the IoT paradigm. iii)
Exerting adaptive control of the EM behavior of planar materials in
a classic sense-input/adapt-output loop. Thus, the HyperSurface
architecture introduces three functionality layers~\cite{AbadalACCESS,pitilakisMETA2018,Liu2018ISCAS,Liaskos:2018:UAS:3289258.3192336,DBLP:conf/WoWMoM/Liaskos}:
the \uline{intra-networking layer}, the \uline{gateway layer}
and the \uline{software control layer}. 

\textbf{\uline{The intra-networking layer}} comprises a set of
networked controllers embedded within the HyperSurface material. One
controller is described as a head node responsible for controlling
N active elements ($1:N$), i.e., relaying actuation directives to
each active element and gathering sensory information via measurements
(e.g., power flow across active elements) or specific-purpose equipment
(e.g., field sensors). Each controller has communication capabilities,
i.e., at least relay information between itself and the gateway layer,
while the most interesting capabilities result when inter-controller
communication is also enabled~\cite{DBLP:conf/iscas/Tasolamprou}. 

We outline the following general options for the physical implementing
of the controller intra-networking layer:

\textbf{No intra-networking} corresponds to the simplest case where
each controller can communicate only with the gateway layer. For instance,
each controller can be directly wired to an IoT gateway. The no intra-networking
approach offers the benefits of \emph{simplicity}\textendash since
the controllers either have no intelligence (e.g., simple shift registers)
or are absent altogether\textendash and \emph{central control} over
the HyperSurface. However, these benefits come at the expense of control
scalability and single-point-of-failure concerns.

\textbf{Wired intra-networking} considers that there exists a wired
medium that interconnects the controllers, as well as some controllers
to the gateway. The wired means can refer to any nature, e.g, copper
wiring, optical fiber and more. Wired networking enables the embedding
of distributed intelligence within the HyperSurface, allowing the
controllers to sense, cooperate and act locally, without the intervention
of the gateway. This adds the benefit of limiting the physical footprint
of the control elements within the HyperSurface, minimizing any degrading
effects to its EM efficiency.

\begin{figure*}[!t]
\centering
\vspace{-0.2cm}
\subfigure[System model\label{fig:system}]{\includegraphics[width=0.65\textwidth]{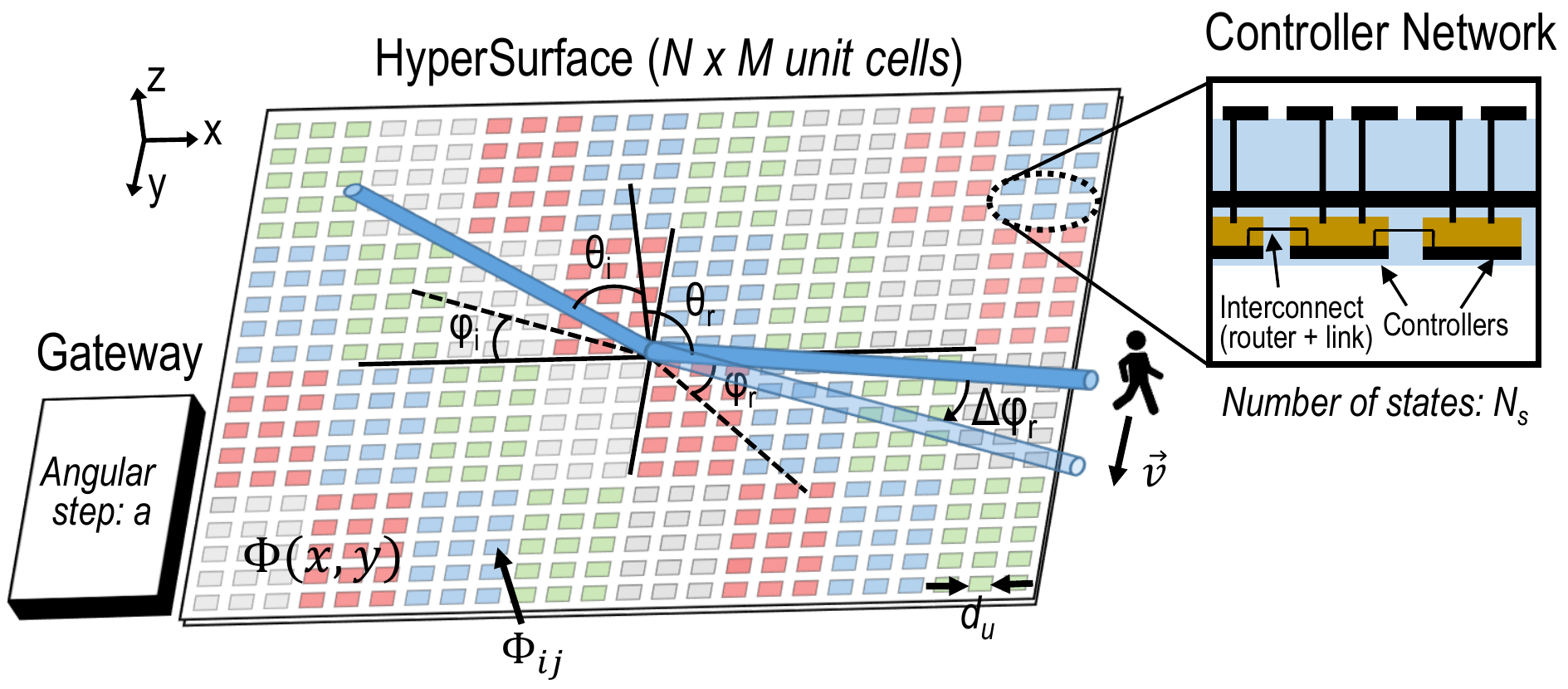}}
\hspace{0.2cm}
\subfigure[Mobility cases (top view)\label{fig:moves}]{\includegraphics[width=0.2\textwidth]{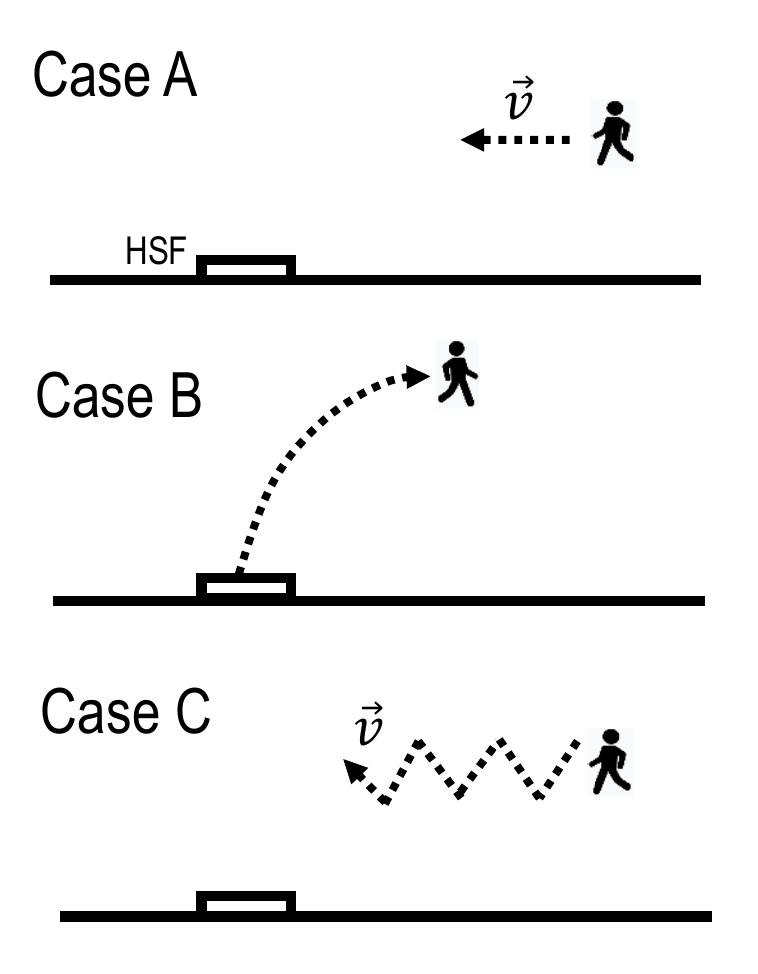}}
\vspace{-0.4cm}
\caption{System model. Target moves with speed $\Vec{v}$ changing the required reflected angle. A HSF with $N\times M$ unit cells of size $d_{u}$ implements a phase profile $\Phi(x,y)$ to obtain the desired reflected angle. The phase of each unit cell $\Phi_{ij}$ is approximated to the closest phase amon the $N_{s}$ available states. The gateway reconfigures $\Phi_{ij}$ when angles vary more than $a$.}
\vspace{-0.2cm}
\label{fig:model}
\end{figure*} 

\textbf{Wireless intra-networking} assumes that each controller is
equipped with a wireless transceiver module, using the very material
of the HyperSurface as the communication channel. The addition of
such transceivers provides the benefits of: i) zero wiring and minimal
parasitic effects the EM efficiency of the HyperSurface, ii) easier
HyperSurface assembly, iii) more densely connected controllers in
general, reducing the network distance between controllers and facilitating
faster intra-communication, and iv) natural data broadcast support,
as opposed to the point-to-point inclination of the wired intra-networking
option. We also highlight the existence of nano-transceivers that
offer minimal complexity and are able to operate in battery-less (energy-harvesting)
mode~\cite{Tabesh.2014,DBLP:conf/camad/TsioliaridouLI18}.

\textbf{\uline{The gateway layer}} constitutes of the hardware
and communication protocols that stand between the HyperSurface 
and any external smart device (mobile phones, IoT devices, etc.).
The gateway can be considered as a two-sided electronic module that
is embedded on the HyperSurface. On the inner-side, it participates
as a peer to the intra-network, being able to address controllers,
sending to and receiving data from them. On the outer-side, it provides
mainstream connectivity by translating the internal data to a form
compatible with common protocols, such as WiFi, Ethernet, Bluetooth,
etc. In this context, we define to variations of the gateway: i)~The
described, \textbf{pass-through mode}, where the gateway acts as translator
of commands between the external world and the intra-controller network,
and ii)~the \textbf{smart mode} where the gateway can perform additional
duties, such as the monitoring of the intra-network for errors and
malfunctions, intra-routing assistance, e.g., via the smart multi-entrypoint
diffusion of data within the intra-network for advanced performance,
and handling computational duties to alleviate the controllers~\cite{saeed2018fault}.

\textbf{\uline{The software control layer}} comprises the software
required for interacting with the HyperSurface while abstracting the underlying physics.
Its role is to make HyperSurfaces easily integrate-able
into applications and systems. To this end, it comprises of~\cite{HSFAPI}:
i)~The \textbf{HyperSurface Programming Interface (HPI)}, which exposes
the EM capabilites of a HyperSurface in a library of software callbacks,
ii)~The \textbf{User-side Interrupt Service}, i.e., a user-side daemon
that receives and handles interrupts, e.g., messages spontaneously
generated by HyperSurfaces to denote hardware failures, and iii)\textbf{~The
EM Compiler}, a HyperSurface-\emph{external} or \emph{-internal} software
service that translates the HPI callbacks to corresponding spatio-temporal
actuation directives for the HyperSurface active elements~\cite{HSFComp}. 

At its most advanced implementation, the HSF paradigm can
advance to a fully autonomic operation, where its controllers can
directly measure and profile the EM cross-interactions between \emph{variable
meta-atoms}, i.e., ones that are in non-periodic layouts and different
in shape, material composition, size, and even mobile. Subsequently,
the Gateway can call upon an internal compilation service to obtain
the EM behavior set by a user via the HPI.

\section{\label{sec:programming}System Model}
This section outlines the notation and main assumptions. Let us consider an HSF with $M\times N$ unit cells and a gateway as shown in Figure \ref{fig:model}. Each unit cell is squared with side $d_{u}$ and can be configured in any of its $N_{s}$ states ($N_{s} = 4$ in Fig. \ref{fig:system}, with each state represented by a color). The metasurface is fully illuminated by a plane wave coming from $\{\theta_{i},\phi_{i}\}$ that is reflected towards $\{\theta_{r},\phi_{r}\}$. The aim of the HSF is to adapt to changes in either direction.

\subsection{Mobility model}
\begin{figure*}[!t]
\centering
  \includegraphics[width=0.9\textwidth]{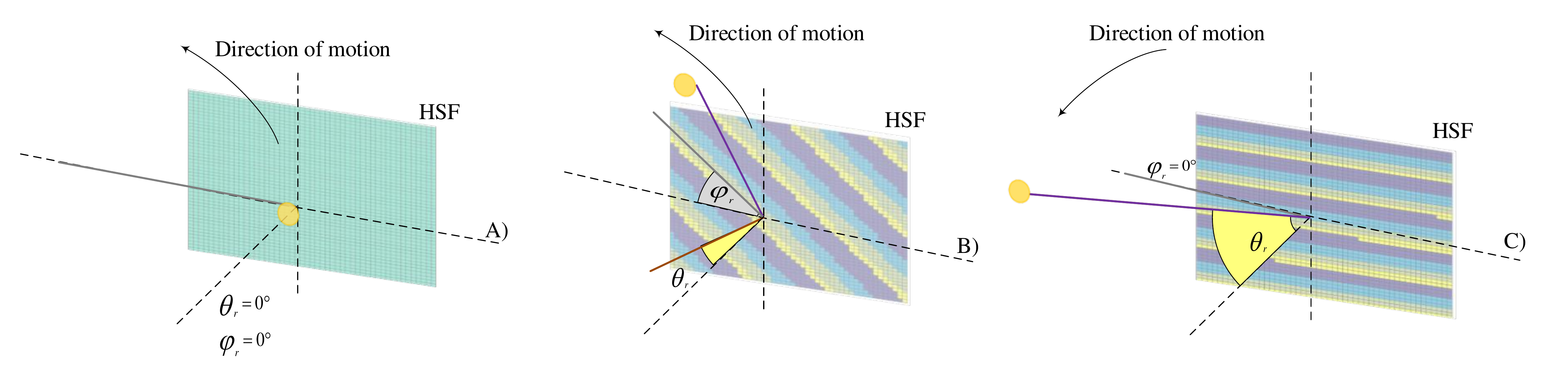}
  \vspace{-0.3cm}
  \caption{Illustration of tracking in Case $B$: projectile movement parallel to the HSF.}
  \vspace{-0.2cm}
  \vspace{-0.2cm}
\label{walk}
\end{figure*}

Without loss of generality, let us consider that the HSF is used to keep track the position of users or objects moving along a vector $\Vec{v}$. A possible application would be to maintain wireless connectivity in 5G with highly-directive antennas \cite{Liaskos2018}. We study three types of trajectories, whose top view is illustrated in Fig. \ref{fig:moves}: 
\begin{itemize} 
    \item \textbf{Case $A$:} the target moves in a straight line parallel to the surface where the target is at the same height than the surface. 
    Motion starts from a point far away from the surface and finishes when the object is directly in front of it. This scenario represents the movement of mobile user, thus we assume a default speed of 1.4 m/s, which is the speed of walking of the average human. 
    \item \textbf{Case $B$:} the target describes a projectile motion parallel to the surface starting from a point close to the surface and moving away from it, as illustrated in Fig. \ref{walk}. This could represent a typical case of radar tracking. Unlike the horizontal movement described above, the projectile motion changes both the azimuth and elevation angles of the reflected signal. The initial speed is assumed to be 30 m/s. 
    \item \textbf{Case $C$:} the target takes arbitrary leaps which results into abrupt changes in location as apposed to the gradual change in the aforementioned cases. This case represents a person moving in an area with multiple mobile obstacles resulting in intermittent connection with the surface. Using the same settings of Case $A$, we model the arbitrary movement by randomly changing the azimuth angle of the reflected signal. 
\end{itemize}

In all cases, we consider that the metasurface is fixed onto a wall and normal incidence. The position of the moving objects can be expressed in spherical coordinates using the metasurface as point of reference. The reflected angles are obtained accordingly.

\subsection{Metasurface Coding}
To achieve anomalous reflection, the metasurface is modeled as a reflectarray with linear phase gradients $\Phi'_{x}$ and $\Phi'_{y}$ in the $x$ and $y$ directions, respectively, as shown in Fig. \ref{fig:system} \cite{Yu2011a}. In such a configuration, the metasurface would ideally present full reflectivity and a phase profile $\Phi(x,y) = \Phi'_{x}x + \Phi'_{y}y $ so that it complies with the required phase gradients. With this phase profile, the momentum conservation law for wave vectors at the HSF interface yields
\begin{equation}\label{eq:dphi}
\begin{array}{l}
k_{i} \sin{\theta_{i}}\cos{\phi_{i}} + \Phi'_{x} = k_{r} \sin{\theta_{r}}\cos{\phi_{r}} \\
k_{i} \sin{\theta_{i}}\sin{\phi_{i}} + \Phi'_{y} = k_{r} \sin{\theta_{r}}\sin{\phi_{r}}
\end{array} 
\end{equation}
where $k_{i} = \tfrac{2\pi}{\lambda_{i}}$ and $k_{r} = \tfrac{2\pi}{\lambda_{r}}$ are the wave vectors of the incident and reflected medium with their respective wavelengths $\lambda_{i}$ and $\lambda_{r}$. Isolating $\Phi'_{x}$ and $\Phi'_{y}$ from Eq. \eqref{eq:dphi}, we can obtain the required phase gradients. Finally, since our assumed metasurface is able to tune the phase at a unit cell granularity, the phase $\Phi_{ij}$ of the element at column $i$ and row $j$ has to be
\begin{equation}\label{eq:phases}
\Phi_{ij} = (\Phi'_{x}i + \Phi'_{y}j)d_{u}.
\end{equation}

Unit cells have a discrete number of states $N_{s}$. Since we are interested in manipulating the phase in the $2\pi$ range, the unit cell is designed so that its $n$-th state yields a phase shift of $\tfrac{2\pi}{n}$ \cite{Cui2014}. Each unit cell will be assigned the state that is closer to the ideal phase profile $\Phi_{ij}$.


\subsection{Gateway and Controllers}
Let us consider a simple HSF where each controller drives a single unit cell. For the purposes of traffic analysis, we will consider that the HSF is equipped with a gateway that:
\begin{itemize}
    \item has external sensing capabilities to obtain the angle of incidence $\{\theta_{i},\phi_{i}\}$,
    \item communicates with other external devices to obtain the targeted direction $\{\theta_{r},\phi_{r}\}$,
    \item can compute the phase profile $\Phi_{ij}$ and, thus, the state of each unit cell using the model above.
\end{itemize}
Whenever the target angle changes by more than a given pre-defined amount, which we define as \emph{angular step} $a$, the gateway computes the new phase profile and sends packets only to those unit cells that need to change its state.



\section{\label{sec:methodology}Evaluation Methodology}
\begin{figure}[!t]
\centering
\includegraphics[width=\columnwidth]{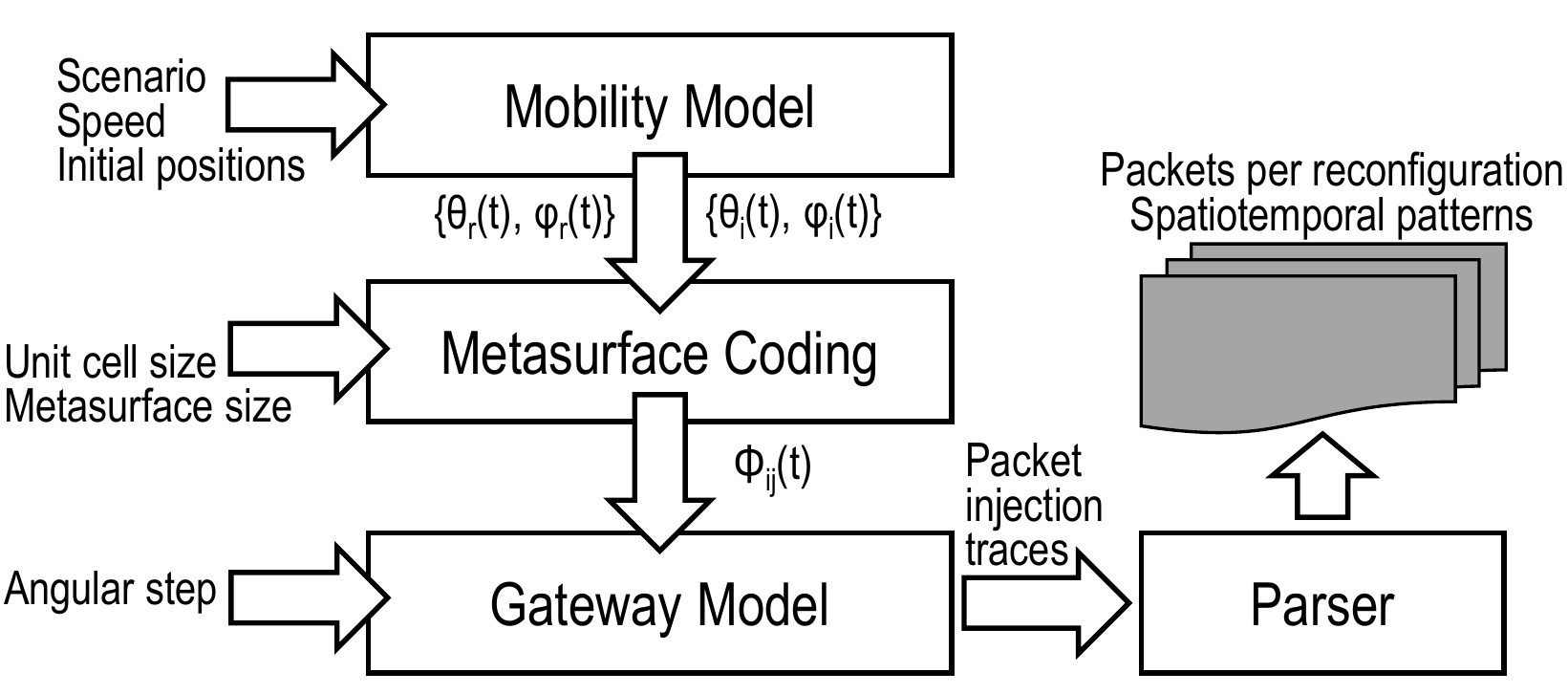}
\vspace{-0.4cm}
\caption{Summary of the evaluation methodology.}
\vspace{-0.2cm}
\label{fig:methodology}
\end{figure}

Figure \ref{fig:methodology} illustrates our methodology to obtain the traffic within the metasurface. MATLAB scripts are employed to simulate the movement and to obtain the corresponding unit cell state matrices. More specifically, we first evaluate the incidence and reflection angles given the positions of the HSF, the illumination source, and the moving target (mobility model). Then, we use Eq. \eqref{eq:dphi} to obtain the phase gradients and then Eq. \eqref{eq:phases} to calculate the phase $\Phi_{ij}$ of each unit cell. Finally, we take the unit cell state that yields the phase that is closer to $\Phi_{ij}$ (metasurface coding). In all cases, we assume that the HSF has $N\times M=50\times 50$ unit cells.

The methods described above allow to obtain the matrix of unit cell states for any incidence and reflection conditions. Iterating over such calculations, we can obtain successive unit cell state matrices corresponding to a given movement with any angular granularity. To model the gateway, we only obtain the unit cell state matrices in steps corresponding to the \emph{angular step} parameter. A \emph{diff} operation between adjacent unit cell state matrices describes which unit cells need to changed and, given our assumptions, which unit cells will receive packets from the gateway. 

For each type of movement, we collect traces that describe the time instant at which packets are generated as well as their intended destinations. The traces are obtained and classified as a function of the following parameters:

\begin{itemize}
\item \textbf{Number of states $N_{s}$:} the calculations used to assign cell states rounds to the state that yields the phase which closer to $\Phi_{ij}$. Higher number of states corresponds to a finer resolution phase gradient that improves the metasurface performance, but possibly at the cost of trasmitting more packets. By default, we set $N_{s}=4$.
\item \textbf{Angular step $a$:} some applications may require tracking objects at very fine-grained angular resolution. In that case, the gateway would probably need to trigger state changes more often. This is modeled via the angular step, whose default value is set to $a=5^{\circ}$.
\end{itemize}

\noindent Finally, we parse the traces to obtain the following relevant metrics:
\begin{itemize}
\item \textbf{Percentage of state-changing cells:} The comparison between successive HSF state matrices determines the percentage of cells that must be adjusted to accommodate the change in the position of the target or the illumination source.  
\item \textbf{Destination matrix:} The traffic analysis yields a matrix containing the ratio of packets delivered to a given destination with respect to all the transmitted messages.
\end{itemize} 


\section{\label{sec:results}Workload Characterization}
\begin{figure*}[!t]
    \centering
    \subfigure[\label{recreq}]{\includegraphics[width=0.3\textwidth]{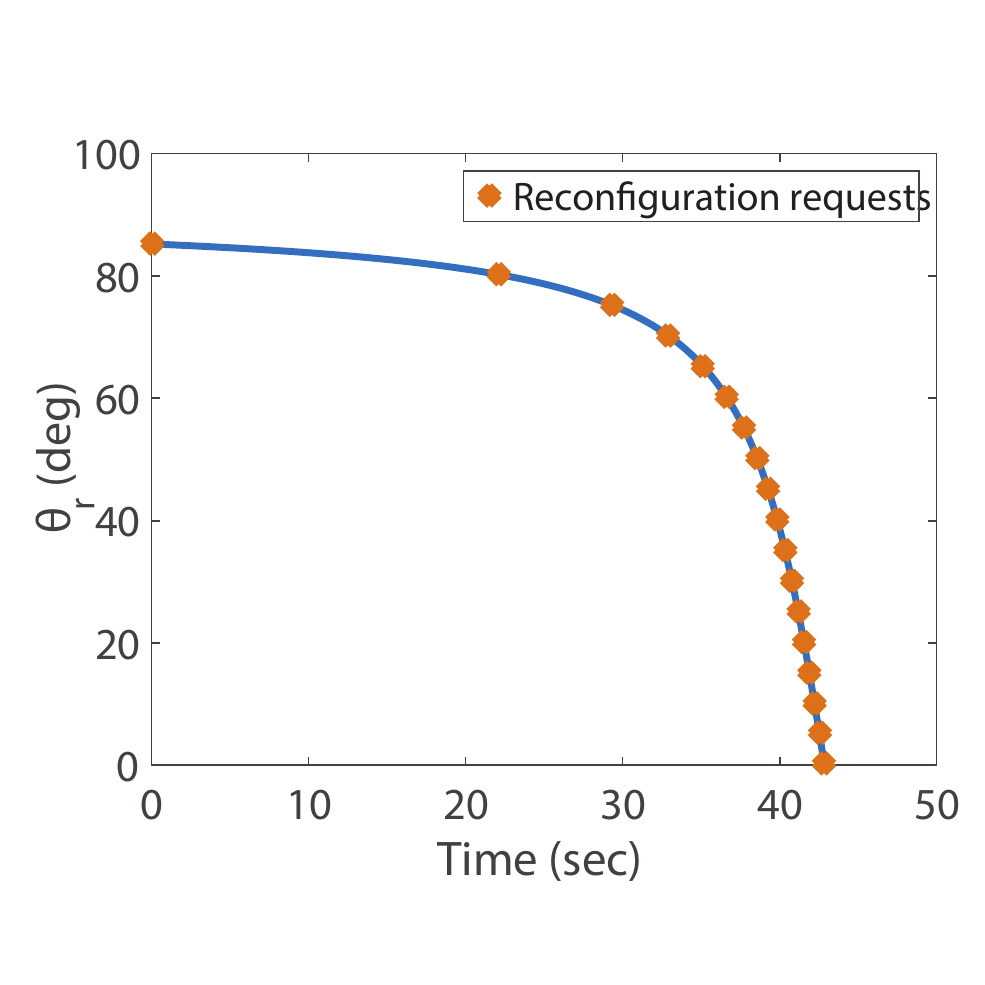}}
    \subfigure[\label{recper}]{\includegraphics[width=0.3\textwidth]{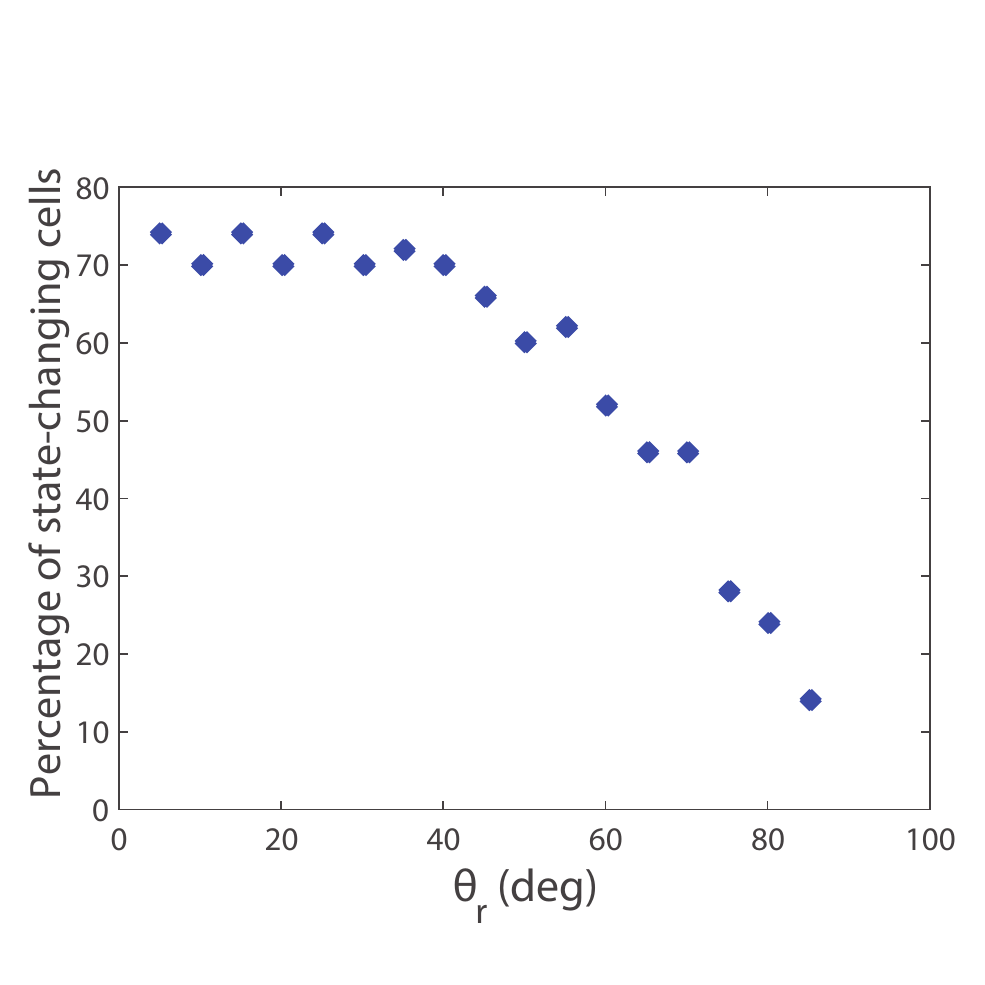}}
    \subfigure[\label{ratstr}]{\includegraphics[width=0.3\textwidth]{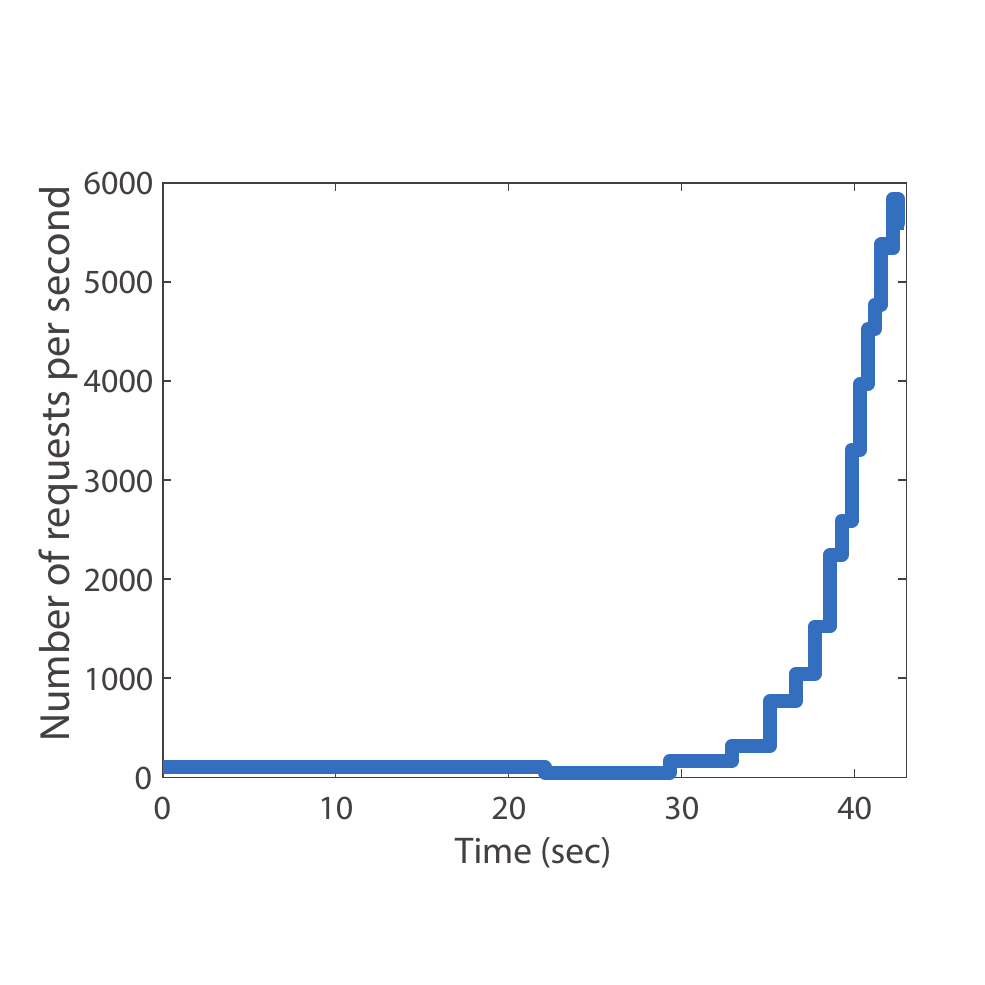}}
    \vspace{-0.5cm}
    \caption{For an object moving according to Case $A$: (a) Reconfiguration requests during the tracking of the object. 
    (b) Percentage of reconfigured cells versus the azimuth angle of the reflected signal. (c) The number of reconfiguration requests per second.}
    \label{both_}
    \vspace{-0.2cm}
\end{figure*}

In this section, we characterize the traffic workload on the controller network. Specifically, we analyze the spatio-temporal intensity and rate of updates generated by different types of movements. This is accomplished 
by showing how often reconfiguration commands are injected to the system. We use this to asses the rate of injection throughout the tracking process. In addition, we visualize the spatial distribution of the required configuration through heat maps that correspond to the surface. 

The generation of reconfiguration requests relies on the location of the tracked object, the motion pattern and the angular step. While the HSF can only sense the change in the incident and reflection angles, the latter is an outcome of the distance and the height difference between the tracked object and the surface, and the motion pattern. This is depicted in Fig. \ref{recreq} where a motion of Case $A$ is tracked. The markers indicate the time instances when reconfiguration requests are sent from the gateway to the system. Fig. \ref{recreq} shows that reconfigurations are more frequent as the object moves closer to the surface. For example, more than $88\%$ of the reconfigurations are required within the last third of the motion. This is expected since the reflection angle changes faster when the object is closer to the surface.   

Note that not all cells are reconfigured at each request. 
From Fig. \ref{recper}, it is clear that the percentage of reconfigured cells is highly dependent on the preceding and the currently targeted angles. For example, when the tracked object moves from $\theta_{r} = 25^{\circ}$ to $\theta_{r} = 20^{\circ}$, $70\%$ of the cells are reconfigured. When the change is from $\theta_{r} = 80^{\circ}$ to $\theta_{r} = 75^{\circ}$, on the other hand, only $28\%$ of the cells are reconfigured in spite of the fact that the change in both cases was $5^{\circ}$. 

When a change in the state of the HSF is required, reconfiguration requests are streamed into the system, which renders the injection inherently bursty (i.e. a burst of reconfiguration commands are injected every time a proper change in the angle occurs). However, the number of reconfiguration requests and the frequency with which the requests are made determine the injection rate. In Fig. \ref{ratstr}, we show how the injection rate significantly increases when the changes in the reflection angle are more frequent. Coincidentally, frequent changes in Case $A$ occur for target angles affecting a higher percentage of unit cells, pushing the injection rate further. This is an important result because excessive injection rates can be an offset of congestion within the controller network.



\begin{figure}[!t]
    \centering
    \subfigure[Case $A$\label{hmb_c1}]{\includegraphics[width=0.32\columnwidth]{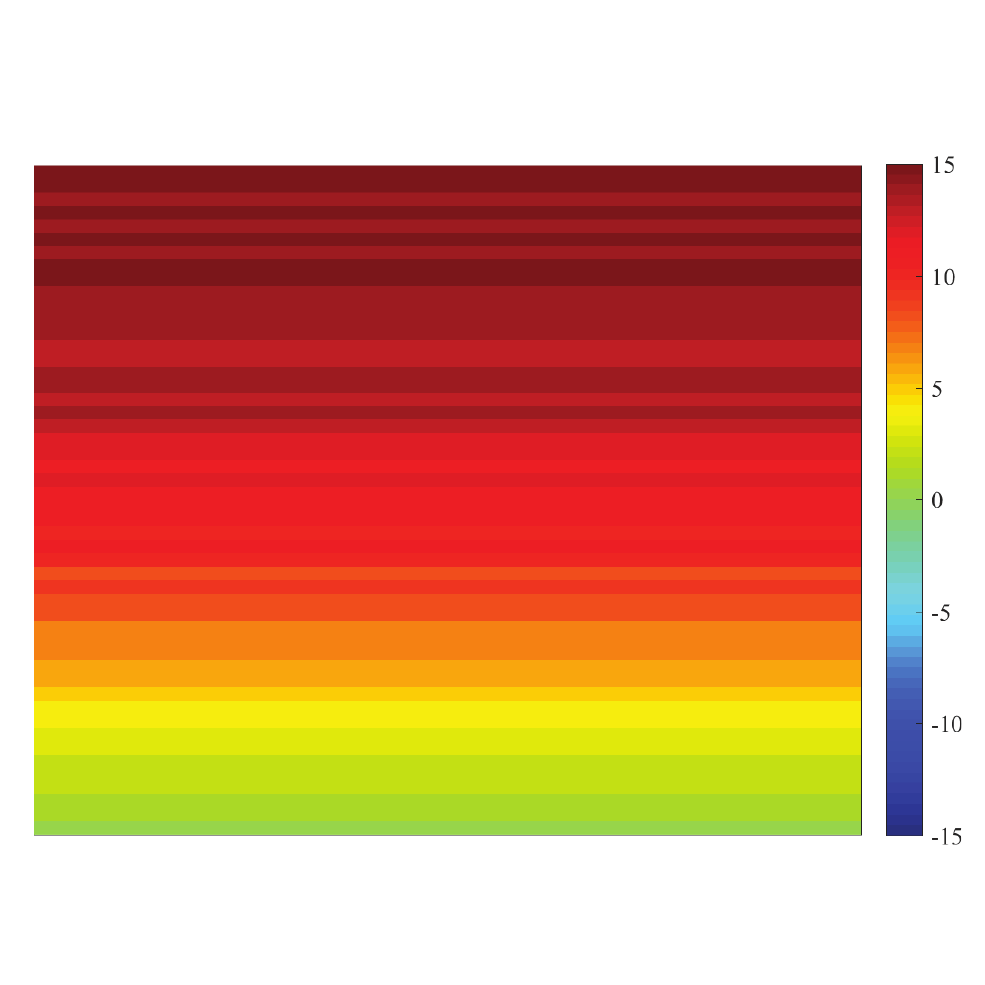}}\vspace{-0.05cm}
    \subfigure[Case $B$\label{hmb_c2}]{\includegraphics[width=0.32\columnwidth]{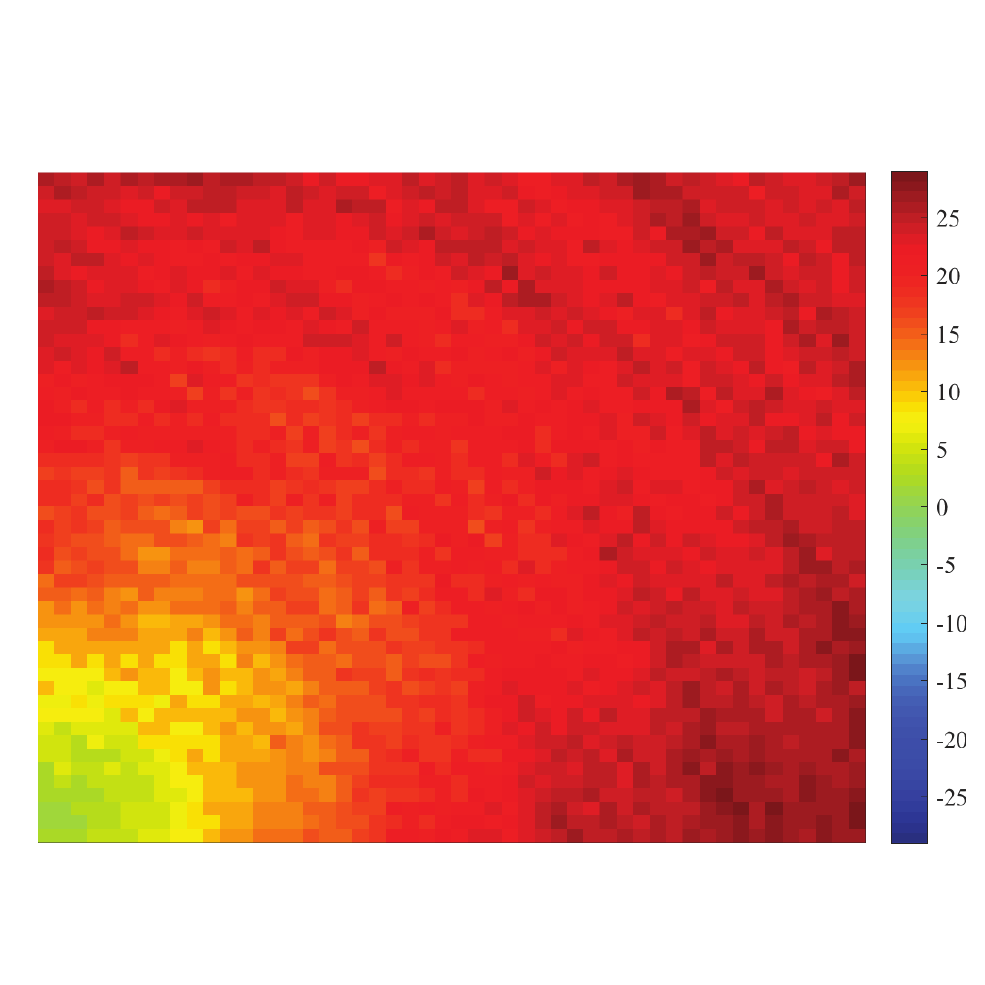}}\vspace{-0.05cm}
    \subfigure[Case $C$\label{hmb_c3}]{\includegraphics[width=0.32\columnwidth]{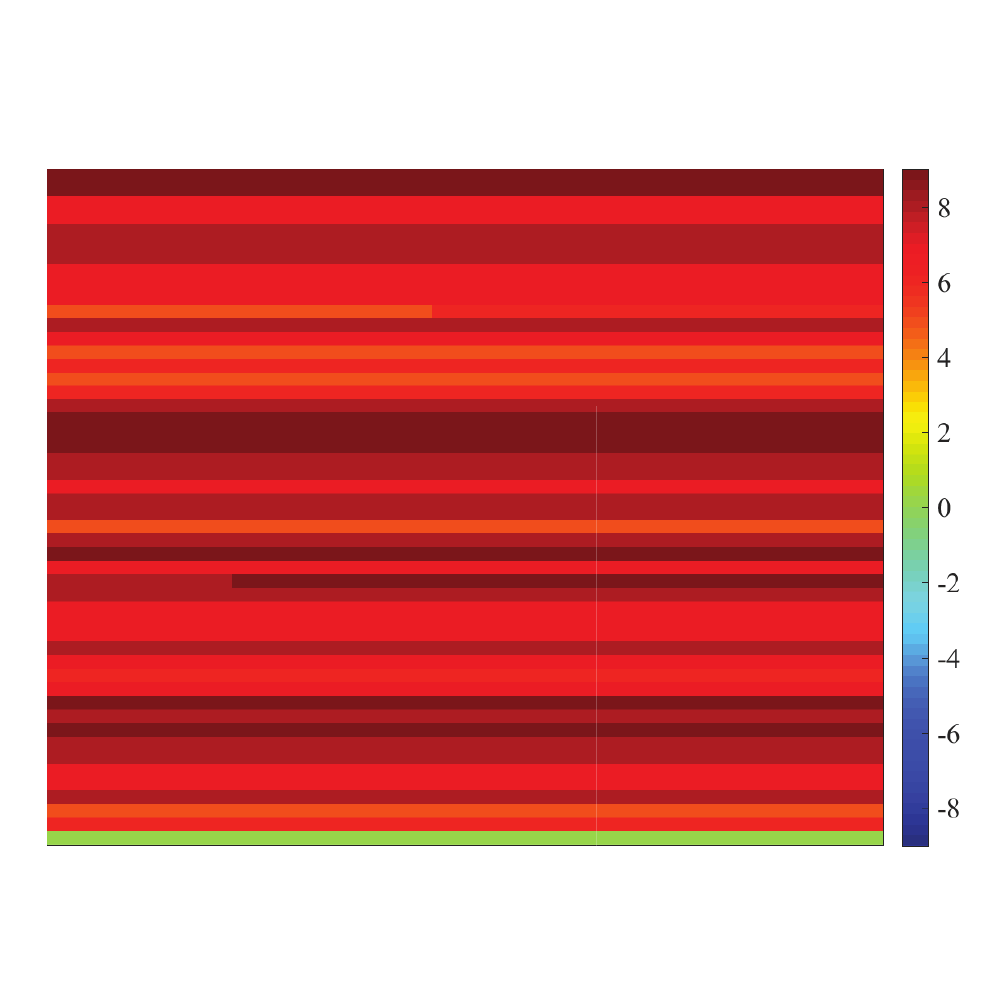}}\vspace{-0.05cm}
    \vspace{-0.3cm}
    \caption{Spatial distribution of traffic for the three considered movement scenarios.}
    \label{heatnap_all}
    \vspace{-0.2cm}
\end{figure}

Another way to evaluate traffic is the spatial distribution of cells to be updated. To visualize this we resort to heat maps, where hotter spots represent the regions of the HSF where controllers receive higher numbers of packets. Fig. \ref{heatnap_all} depicts the heat maps corresponding to the three scenarios of movements. Fig. \ref{hmb_c1} shows the case of gradual change in $\theta_{r}$ where the tracked object follows the motion of Case $A$. Fig. \ref{hmb_c3}, however, shows the case of arbitrary changes in $\theta_r$ where the object makes sudden unpredictable leaps, i.e. Case $C$. The heat maps demonstrate that the traffic is almost evenly distributed over the surface for arbitrary changes. In addition, if compared with the heat maps in Fig. \ref{hmb_c2}, one can observe the difference in the traffic distribution when the elevation angle of the reflected wave is fixed to $0$ (i.e. the HSF is at the same height of the tracked object) and when it is variable. This information can be used in designing the routing mechanism, congestion control techniques and in placing the HSF tiles.

We then change some of the metasurface parameters, namely the angular step and the number of states and observe the effects on the performance. In the previous results, the angular step was set to $5^{\circ}$. The value of $a$ can be used as an indication of the beam width, such that smaller values of $a$ infer narrower beam widths and higher tracking precision requirements. We investigate the impact of changing $a$ when an object moving according to Case $B$ is tracked. Case $B$ implies that both the azimuth and elevation angles are varied throughout the movement. A reconfiguration is requested every time either of the two angles change by $a^{\circ}$.

\begin{figure}[!t]
\centering
\vspace{-0.2cm}
\includegraphics[height=5 cm]{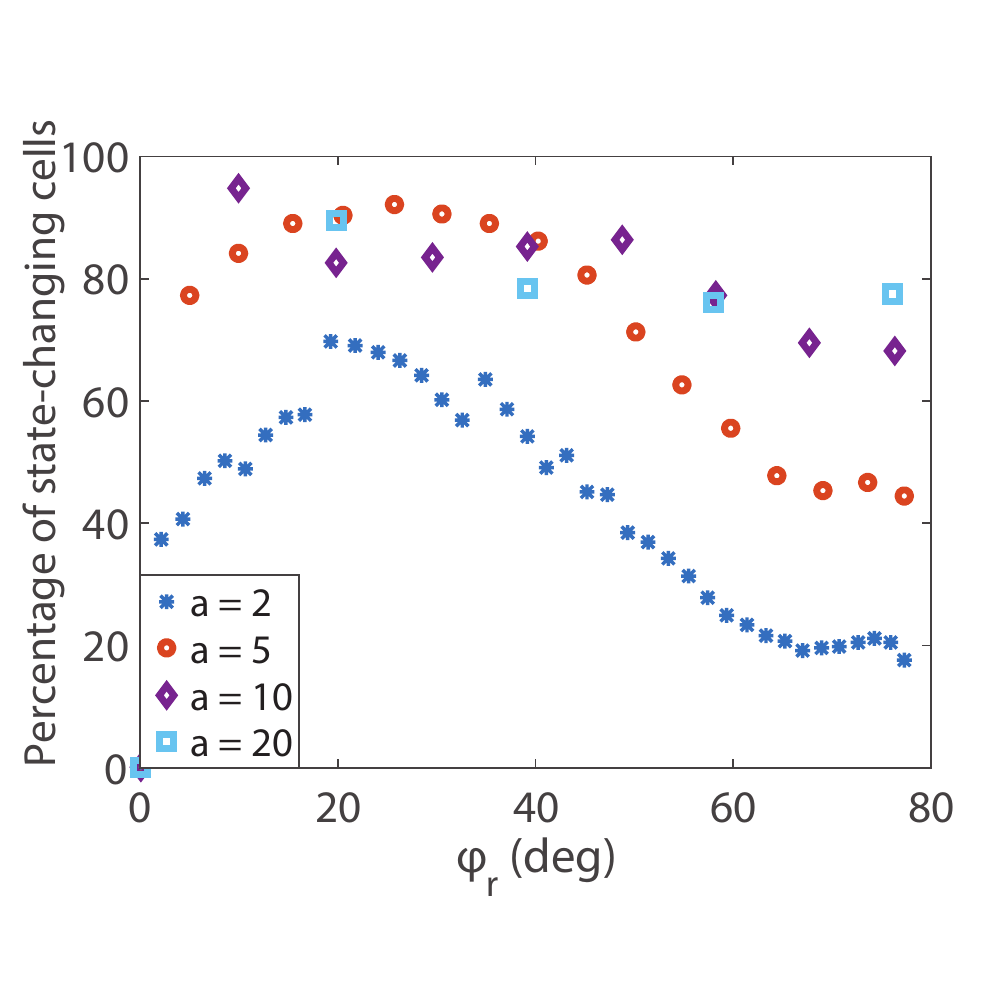}
\vspace{-0.2cm}
\caption{Elevation angle of the reflected signal versus the percentage of state-changing cells for different values of the angular step for projectile motion.}
\vspace{-0.2cm}
\label{angstep}
\end{figure}

Fig. \ref{angstep} depicts the percentage of state-changing cells corresponding to different values of the elevation angle of the reflected signal for different values of $a$. An angular step of $2^{\circ}$ achieves the smallest percentage of change, and thus traffic, over the entire range of angle variation, which suggests that the HSF may operate faster. However, small angular step requires a narrow beam which might increase the complexity of the surface fabrication. In addition, as the value of $a$ decreases the rate of updates is expected to increase. The effects of this increase on the injection rate will be investigated in the future. It is worth mentioning here that the movement in Fig. \ref{angstep} is projectile movement. However, for clarity we only show the change in angles in the first half of the motion (until the object reaches the highest point and before dropping).  

\begin{figure}[!t]
\centering
\vspace{-0.2cm}
\includegraphics[width= \columnwidth]{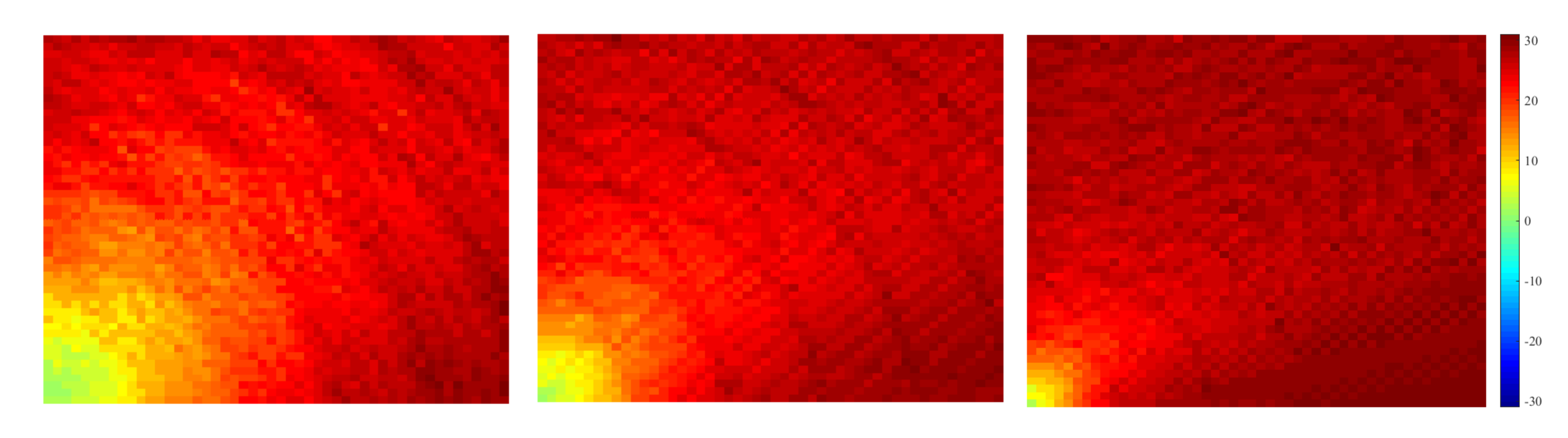}
\vspace{-0.6cm}
\caption{Spatial distribution of traffic in the case of projectile motion for values of $N_{s}$ of $N_{s}=4$ (left), $N_{s}=8$ (middle) and $N_{s}=16$ (right).}
\vspace{-0.2cm}
\label{heatnap_b}
\end{figure}

Fig. \ref{heatnap_b}, on the other hand, shows the heat maps produced from tracking an object in the projectile movement for different numbers of states of the HSF cells, namely $N_{s} = 4,8,16$. As the number of states increases, the traffic becomes heavier over the entire surface, while for smaller number of states the traffic is lighter especially at the bottom-left corner. This might stem from the fact that the bottom-left corner is where the programming starts (initial state is always zero phase). In rather large phase gradients, that area remains largely at the same state.

\section{\label{sec:conc}Conclusion}
This paper has proposed and tested a comprehensive methodology for the characterization of the communication requirements of HSFs. We have analyzed the beam steering case for different models of moving targets and varying multiple design parameters. It has been observed that traffic is inherently bursty with an uneven spatial distribution and that the load increases significantly as objects approach the HSF and for HSFs with a larger number of unit cell states. Moreover, there exists a tradeoff between the rate of updates and the amount of messages per update that is regulated by the angular granularity of the tracking: the results suggest that finer tracking may result in higher yet less bursty load. In future works, we expect to use the obtained communication traces to perform a preliminary evaluation of intra-HSF networks, aiming to determine the reconfiguration capabilities of future HSFs.

\begin{acks}
This work was supported by the European Union's Horizon 2020 research and innovation programme-Future Emerging Topics (FETOPEN) under grant agreement No 736876.
\end{acks}

%
\bibliographystyle{ACM-Reference-Format}
\bibliography{ref-short.bib}

\end{document}